\documentstyle[onecolumn]{mn}

\newcommand{\mincir}{\raise -2.truept\hbox{\rlap{\hbox{$\sim$}}\raise5.truept
\hbox{$<$}\ }}
\newcommand{\magcir}{\raise -2.truept\hbox{\rlap{\hbox{$\sim$}}\raise5.truept
\hbox{$>$}\ }}
\newcommand{\minmag}{\raise-2.truept\hbox{\rlap{\hbox{$<$}}\raise 6.truept\hbox
{$>$}\ }}
\newcommand{\be}{\begin{equation}}
\newcommand{\ee}{\end{equation}}
\newcommand{\ba}{\begin{eqnarray}}
\newcommand{\ea}{\end{eqnarray}}
\newcommand{\brr}{\begin{array}}
\newcommand{\err}{\end{array}}
\newcommand{\bc}{\begin{center}}
\newcommand{\ec}{\end{center}}

\title{Velocity fields in non--Gaussian cold dark matter models}
\author[F. Lucchin et al.]
{F. Lucchin$^{1}$, S. Matarrese$^{2}$,
A. Messina$^{3}$, L. Moscardini$^{1}$ and G. Tormen$^{4}$
\thanks{Present address: Institut d'Astrophysique de Paris,
98 bis Boulevard Arago, F--75014 Paris, France}\\
$^{1}$ Dipartimento di Astronomia, Universit\`{a} di Padova, vicolo
dell'Osservatorio 5, I--35122 Padova, Italy\\
$^{2}$ Dipartimento di Fisica {\em Galileo Galilei}, Universit\`{a} di
Padova, via Marzolo 8, I--35131 Padova, Italy\\
$^{3}$ Dipartimento di Fisica {\em Augusto Righi}, Universit\`{a} di
Bologna, via Irnerio 46, I--40126 Bologna, Italy\\
$^{4}$ Institute of Astronomy, University of Cambridge,
Madingley Road, Cambridge CB3 0HA, UK}

\begin{document}

\maketitle

\begin{abstract}
We analyse the large--scale velocity field obtained by N--body simulations of
cold dark matter (CDM) models with non--Gaussian primordial density
fluctuations, considering models with both positive and negative primordial
skewness in the density fluctuation distribution. We study the velocity
probability distribution and calculate the dependence of the bulk flow,
one--point velocity dispersion and Cosmic Mach Number on the filtering size. We
find that the sign of the primordial skewness of the density field provides
poor discriminatory power on the evolved velocity field. All non--Gaussian
models here considered tend to have lower velocity dispersion and bulk flow
than the standard Gaussian CDM model, while the Cosmic Mach Number turns out to
be a poor statistic in characterizing the models. Next, we compare the
large--scale velocity field of a composite sample of optically selected
galaxies as described by the Local Group properties, bulk flow, velocity
correlation function and Cosmic Mach Number, with the velocity field of mock
catalogues extracted from the N--body simulations. The comparison does not
clearly permit to single out a best model: the standard Gaussian model is
however marginally preferred by the maximum likelihood analysis.
\end{abstract}

\begin{keywords}
Cosmology: theory -- dark matter -- galaxies: clustering, formation --
large--scale structure of Universe -- early Universe
\end{keywords}

\section{Introduction}
The cornerstone of current theories of structure formation in the universe is
the dominance of a non--baryonic dark matter component. During the last decade
the standard cold dark matter (hereafter SCDM) scenario has shown a high
predictive power in explaining many observed properties of the large--scale
galaxy distribution: the constituents of dark matter in this model are massive
particles, which decoupled when non relativistic or have never been in thermal
equilibrium; the primordial perturbations are assumed to be Gaussian and
adiabatic with a scale--invariant power--spectrum, $P(k) \propto k^n$, with
$n=1$ (the so--called Harrison--Zel'dovich spectral index); the SCDM scenario
is also characterized by an Einstein--de Sitter universe with vanishing
cosmological constant. The amplitude of the primordial perturbations is usually
parametrised by the linear {\it bias} factor $b$, defined as the inverse of the
{\it rms} mass fluctuation on a sharp--edged sphere of radius $8~h^{-1}$ Mpc
(in this work we will adopt the value $h=0.5$ for the Hubble constant $H_0$ in
units of $100$ km ${\rm s}^{-1} {\rm Mpc}^{-1}$). The {\it COBE} DMR detection
of large angular scale anisotropies of the cosmic microwave background (Smoot
et al 1992; Bennett et al. 1994) can be used to fix the normalization of the
model, leading to $b \approx 1$. This normalization makes the model completely
specified.

However, it is well known that this model presents some serious problems,
mostly due to the high ratio of small to large--scale power: in particular, the
{\it COBE} normalization implies excessive velocity dispersion on Mpc scale
(e.g. Gelb \& Bertschinger 1994) and is unable to reproduce the slope of the
galaxy angular correlation function obtained from the APM survey (Maddox et al.
1990).

To overcome these difficulties many alternatives to this basic model have been
proposed: $i)$ ``tilted" (i.e. $n < 1$) CDM models (Vittorio, Matarrese \&
Lucchin 1988; Adams et al. 1992; Cen et al. 1992; Lucchin, Matarrese \&
Mollerach 1992; Tormen et al. 1993, Moscardini et al. 1994), $ii)$ hybrid (i.e.
hot plus cold) dark matter models (e.g. Klypin et al. 1993, and references
therein), $iii)$ CDM models with a relic cosmological constant (e.g.
Efstathiou, Bond \& White 1992, and references therein), $iv)$ non--Gaussian
CDM models (hereafter NGCDM; Moscardini et al. 1991; Messina et al. 1992).

In the present work we consider the last alternative. Physical motivations for
this class of models can be given in terms of the effect of relic topological
defects (e.g. Scherrer \& Bertschinger 1991; Scherrer 1992, and references
therein), or in terms of the inflationary dynamics of models containing
multiple scalar fields (see e.g. Salopek 1992), or in the frame of the cosmic
explosion scenario (e.g. Weinberg, Ostriker \& Dekel 1989). NGCDM models have
been investigated in a series of papers (Moscardini et al. 1991; Matarrese et
al. 1992; Messina et al. 1992; Coles et al. 1993a,b; Moscardini et al. 1993;
Borgani et al. 1994), mainly devoted to the analysis of the clustering
properties of the matter distribution on large scales, as resulting from
N--body simulations. A similar analysis was done by Weinberg \& Cole (1992),
who, however, performed numerical simulations with scale--free initial
power--spectra. Of course, the ultimate probe of the non--Gaussian character of
the primordial perturbation field can only be obtained from the analysis of the
cosmic microwave background (CMB) anisotropies on large angular scales: a
number of statistical tests have been recently proposed in order to detect
possible non--Gaussian signatures, such as the skewness of the temperature
distribution (Hinshaw et al. 1994) and the genus of iso--temperature contours
(Smoot et al. 1994). The analyses performed on the first year {\em COBE} data
have revealed that non--Gaussian signals are not present on the angular scales
probed by the DMR experiment ($\magcir 7^\circ$), beyond those due to the
effects of the cosmic variance (e.g. Scaramella \& Vittorio 1991). The
implication of this result is that non--Gaussian features cannot be relevant
for the large--scale gravitational potential, so, either the primordial
fluctuations were indeed Gaussian, or primordial non--random phases were only
present on scales below $7^\circ$, as it seems the case for anisotropies
generated by topological defects (e.g. Coulson et al. 1994, and references
therein). Therefore, for the models we consider here, we just have to require
that the gravitational potential was significantly non--Gaussian already at
redshifts of order of a tenth, when we start to evolve our system, and up to
scales as large as $\sim 10^2$ Mpc (much below the {\em COBE} scale), as probed
by our simulations.

This work is devoted to a detailed study of the velocity field in N--body
simulations of NGCDM models, and to a comparison of the large--scale velocity
field of a composite sample of optically selected galaxies (1184 galaxies, with
known radial peculiar velocities, grouped in 704 objects, from the ``Mark II"
compilation), with that obtained from the simulations. In particular, we
calculate the probability to reproduce the observed properties of the Local
Group and the observed values for the bulk flow, the velocity correlation
function, and the Cosmic Mach Number of the data in mock catalogues extracted
from our simulations. A similar analysis has been performed by Tormen et al.
(1993) and Moscardini et al. (1994) in the frame of open and/or tilted CDM
models. Contrary to most previous analyses of NGCDM models (Moscardini et al.
1991; Messina et al. 1992), we here fix the `present time' of the simulations
in such a way that the linear bias parameter $b$ is one, consistently with the
{\em COBE} normalization\footnote{Even though the statistical analysis of CMB
anisotropies on large angular scales for non--Gaussian models cannot be reduced
to calculating the {\em rms} fluctuation we assume here that the effect of
non--Gaussian statistics on the {\em COBE} DMR scale is small, so that we can
safely use the standard normalization, leading to $b\approx 1$.}. The same
choice for the normalization has been made by Moscardini et al. (1993), where
the effect of primordial non--random phases on the large--scale behaviour of
the galaxy angular two--point function has been investigated: it was found that
models with initially negatively skewed fluctuations are in principle capable
of reconciling the lack of large--scale power of the CDM spectrum with the
observed clustering of APM galaxies.

The plan of the paper is as follows. In Section 2 we introduce our skewed CDM
models. In Section 3 we discuss the general properties of the velocity field of
the simulations, analysing the bulk flow, the one--point velocity dispersion
and the Cosmic Mach Number at different smoothing scales. We also analyse the
density and velocity probability distributions and power--spectra. Section 4 is
instead devoted to the statistical comparison of mock catalogues extracted from
the simulations with observational data. Conclusions are drawn in Section 5.

\section{Skewed CDM models and N--body simulations}

The non--Gaussian statistics considered here are those adopted by Moscardini et
al. (1991), namely the {\em Lognormal} (hereafter $LN$) and the {\em
Chi--squared} with one degree of freedom (hereafter $\chi^2$), chosen as
distributions for the peculiar gravitational potential, $\Phi({\bf x})$, before
the modulation by the CDM transfer function. These distributions actually split
in two different types of models: the {\em positive} ($LN_p$ and $\chi^2_p$)
and {\em negative} ($LN_n$ and $\chi^2_n$) models, according to whether the
sign of the skewness for linear mass fluctuations, $\langle \delta_M^3\rangle$,
is positive or negative respectively.

All our model distributions are set up in such a way that $\Phi$ has the
``standard" CDM power--spectrum
$$
{\cal P}_\Phi(k) = {9 \over 4} {\cal P}_0 k^{-3} T^2(k) =
{9 \over 4} {\cal P}_0 k^{-3} [1 + 6.8 k + 72.0 k^{3/2} + 16.0 k^2 ]^{-2},
\eqno(2.1)
$$
where ${\cal P}_0$ is a suitable normalization constant and the CDM transfer
function $T(k)$ is taken from Davis et al. (1985) for a flat universe. This
choice for the spectrum allows us to make a direct comparison with the Gaussian
CDM (hereafter $G$) model.

We use a particle--mesh code with $N_p = 128^3$ particles on $N_g=128^3$
grid--points. The box--size is $L=260 ~h^{-1}$ Mpc. We run two independent
simulations for each of the five considered models.

The primordial gravitational potential is obtained by the convolution of a real
function $\tau({\bf x})$ with a random field $\varphi({\bf x})$,
$$
\Phi({\bf x}) = \int d^3 y \ \tau({\bf y} - {\bf x}) \varphi({\bf y}).
\eqno(2.2)
$$
The field $\varphi$ is obtained by a non--linear transformation on a zero--mean
Gaussian process $w$, with unit variance and power--spectrum $\propto k^{-3}$;
the function $\tau$ is fixed by its Fourier transform,
$$
\tilde{\tau} ({\bf k}) \equiv \int d^3 x e^{-i {\bf k}
\cdot {\bf x}} \tau({\bf x}) = T(k) F(k),
\eqno(2.3)
$$
where $F(k)$ is a positive correction factor which we applied to get the exact
CDM initial power--spectrum of Eq.(2.1). The non--linear transformations from
$w$ to $\varphi$ is $\varphi({\bf x}) \propto e^{w({\bf x})}$ and $\varphi({\bf
x}) \propto w^2({\bf x})$ for $LN$ and $\chi^2$ respectively (Coles \& Barrow
1987; Coles \& Jones 1991; Moscardini et al. 1991).

The clustering dynamics and the present large--scale structure have been shown
(Moscardini et al. 1991) to strongly depend upon the sign of the primordial
skewness: positive models rapidly cluster to a lumpy structure with small
coherence length, while negative models build up a cellular structure by the
slow merging of shells around primordial voids.

\section{General properties of velocity fields}

The velocity field of NGCDM models has not received as detailed a study as its
Gaussian counterpart; a preliminary analysis was done by Moscardini et al.
(1991). For this reason, before comparing the observed cosmic velocity field
with the prediction of our models, we would like to briefly discuss the role of
primordial density skewness on velocity field statistics. In this section we
will compute some general statistics in the idealized situation of a perfect
knowledge of the three--dimensional velocity field. We will carry out our
analysis on the four skewed models (positive and negative $LN$ and $\chi^2$),
plus the Gaussian model. For each model we will analyse the velocity field
defined on a $128^3$ mesh. The field is obtained using the same procedure as in
Kofman et al. (1994). First we interpolated mass and momentum from the particle
distribution on a regular cubic grid by means of a Triangular Shaped Cloud
(TSC) scheme (see Hockney \& Eastwood 1981). We then smoothed the result using
a further Gaussian filter, to define the velocity field also in underdense or
empty regions of the simulations. The Gaussian window had a radius of $275$ km
s$^{-1}$, close to the dynamical resolution of the simulations. Finally, the
velocity at a grid--point was defined as the local ratio of momentum and mass
density.

Since the normalization criteria of the simulations here analysed differ from
those used in the previous papers, we will also present the density
distribution function and power--spectrum as auxiliary tools for our analysis.

Due to the fact that the velocity field is quite sensitive to long wavelength
fluctuations, following Strauss, Cen \& Ostriker (1993), we included in our
simulations the effect of waves larger than the box size. To do so we added to
the velocity of each particle the linear {\em rms} bulk velocity of a cube of
260 h$^{-1}$ Mpc, the size of our simulations. The direction of the bulk
velocity was chosen at random for each simulation. The value of such a bulk
flow, approximately 220 km s$^{-1}$, was calculated for the Gaussian CDM model;
however, being a linear correction, we could safely apply it also to the
non--Gaussian simulations.

In Figures 1a and 1b we plot the projected particle positions and the smoothed
peculiar velocity taken from a slice of depth $4~h^{-1}$ Mpc, for the Lognormal
and Chi--squared models respectively. A comparison with a similar plot for the
Gaussian model is possible looking to Figure 1b in Moscardini et al. (1994),
where the same scale and normalization for the velocity field was used. Even if
the times here considered are different, a first glance confirms previous
results. The longer time evolution amplifies the properties of skew--positive
models: a lumpy structure, with isolated knots surrounded by large regions of
quasi--uniform density. On the contrary, the shorter evolution does not
completely prevent the skew--negative models from displaying their cellular
structure with long filaments, extended sheets and large voids.

\begin{figure}
\centering
\vspace{12cm}
\caption{a) Slices with thickness $4 ~h^{-1}$ Mpc for the skew--negative (top
row) and skew--positive Lognormal models (bottom row). Left column: projected
particle positions. Right column: projected  peculiar velocity field after
smoothing by a Gaussian filter with width $275$ km s$^{-1}$.}
\end{figure}

\begin{figure}
\centering
\vspace{12cm}
\caption*{b) Slices with thickness $4 ~h^{-1}$ Mpc for the skew--negative (top
row) and  skew--positive Chi--squared models (bottom row). Left column:
projected particle positions. Right column: projected  peculiar velocity field
after smoothing by a Gaussian filter with width $275$ km s$^{-1}$.}
\end{figure}

\subsection{Density and Velocity Probability Distributions}

Figure 2 shows for our models the density fluctuation probability distribution
(left panel), its power--spectrum (central panel) and the velocity modulus $v
\equiv |\vec v|$ distribution function (right panel).  In Table 1 we report
also the first and second moment of the velocity distributions and the second
moment of the density fluctuation distributions. One can see that time
evolution has preserved the memory of the primordial skewness of the density
field in all simulations. In fact, negatively skewed models show a larger
number of voids and underdense regions, and develop fewer non--linear
overdensities than the positively skewed models do. The bump shown in the
density histogram at $0 \mincir \delta \mincir 1$ points out the action of
gravity on the initial underdensities. The Gaussian model exhibits an
intermediate behaviour, with a larger number of moderate overdensities but a
shorter tail on the right. This is also confirmed by the values listed in Table
1.

The density power--spectra are given in the second panel of Figure 2. The fast
decay of the curves at high $k$ is due to the smoothing introduced by the TSC
interpolation used to define the density field on a regular cubic grid. The
primordial phase correlation present in the skewed models has caused
non--linear effects also on the scale of the box. In fact, in the
skew--positive models some power has been transferred from long to short
wavelengths; this is another way to say that these models have more non--linear
structure at small scales. The skew--negative ones show instead a general lower
amplitude for the power--spectrum: the ``cross talk" between modes has slowed
down the growth rate for $P(k)$ at all scales. Recalling that velocities are
sensitive to large scale power and looking at $P(k)$ we may expect to find a
higher mean velocity for the Gaussian model, longer tails of very high
velocities (due to small scale non--linear structures) for the positively
skewed models and in general lower velocities for the negatively skewed models.

The velocity distributions are in fact different from one another. The Gaussian
model has the largest {\em rms} velocity of all, whereas the skew--positive
ones have a larger number of very high velocities. The difference is
statistically significant: after our smoothing, in the skew--positive models
there are, for example, nearly 2000 grid--points with a velocity greater than
2000 km s$^{-1}$, whereas the Gaussian distribution has no grid point with such
a high velocity. As for the negative models, while the $\chi^2_n$ indeed has
low velocities (it is the model with the least non--linear structure of all),
the $LN_n$ exhibits a singular behaviour, with a tail of high velocities as the
positive models have: a possible explanation of this tail is the gravitational
``push'' from voids and underdense regions, which are larger and more extreme
in this model than in the $\chi^2_n$.

\begin{figure}
\centering
\vspace{9cm}
\caption{Left panel: the density fluctuation probability distribution. Central
panel: the density power--spectrum. Right panel: the probability distribution
for the velocity modulus $|\bf v|$. The $G$, $LN_n$ $LN_p$ $\chi^2_n$ and
$\chi^2_p$ models are represented by solid, dotted, short--dashed, long--dashed
and dot--dashed lines respectively. In the central panel only, the heavy solid
line refers to the linear prediction.}
\end{figure}

\begin{table}
\centering
\caption{Moments of Probability Distribution Functions.}
\begin{tabular}{lccc}
Model & $\sigma_\delta$ & $\langle \ |\vec v| \ \rangle$  & $\sigma_v$ \\
      &                 & km s$^{-1}$ & km s$^{-1}$ \\
$G$       & 1.15 & 637 & 260 \\
$LN_n$    & 0.76 & 536 & 310 \\
$LN_p$    & 1.26 & 461 & 279 \\
$\chi^2_n$& 0.79 & 562 & 274 \\
$\chi^2_p$& 1.34 & 474 & 275 \\
\end{tabular}
\end{table}

\subsection{Bulk flow, one--point velocity dispersion and Cosmic Mach Number}

In order to further characterize the velocity field of each model, we computed
the bulk flow, the mean residual velocity and their ratio (the so called Cosmic
Mach Number) in top--hat spheres of different radii. We defined these
statistics as follows. The bulk flow is the amplitude of the centre of mass
velocity of the sphere: $v_{bulk}(R) = \left|\sum_{i=1}^n \vec v_i\right|/n$,
where the sum extends to the $n$ grid--points falling within a distance $R$
from the centre of the sphere; the one--point velocity dispersion $\sigma_v$ of
the sphere is the {\em rms} velocity measured in the reference system comoving
with the bulk flow: $\sigma_v^2(R) = \sum_{i=1}^n (\vec v_i - \vec
v_{bulk})^2/n$, where the sum is extended to the same grid--points; the Cosmic
Mach Number is the ratio between the two: ${\cal M}(R) = v_{bulk}(R) /
\sigma_v(R)$.

In particular, for each model we randomly selected 100 different grid--points
from the simulations, and calculated the mentioned statistics in top--hat
spheres centred on them, for a radius of the sphere ranging from 5 $h^{-1}$ Mpc
to 130 $h^{-1}$ Mpc. Figure 3 shows the mean values obtained by averaging the
results from the 100 estimates. For clarity we prefer not to plot the error
bars: they are always less than 5\% for all statistics and different models.

The bulk flow is plotted in the left panel and the velocity dispersion in the
central one. Note that the saturation of the bulk flow at large scales is due
to the random velocity we added to each particle in order to include the effect
of wavelength larger than the box--size, as discussed above. In both cases the
different models show a behaviour consistent with the velocity distribution
function and density power--spectra. The Gaussian model, with more power on
large scales and a higher {\em rms} total velocity (shown in Table 1), has the
highest figures for both the bulk flow and the residual velocities at all
scales. Skew--positive models, which have larger velocity tails but less power
on the largest scales and lower {\em rms} velocities, show the lowest values
for both the bulk flow and the velocity dispersion. Finally, the skew--negative
models, with an intermediate {\em rms} velocity, also have intermediate values
for the bulk flows and residual velocity.

The similar trends for the bulk flow and velocity dispersion in the different
models implies a wash up of most differences when looking at the Cosmic Mach
Number (right panel of Figure 3), except perhaps at the smallest scales. The
five curves in the panel are in fact much closer to each other than any
observational estimate with the data available today could possibly
distinguish. We do not expect that adding the observational uncertainties to
the simulated data (see the next section) can in any way disentangle this
situation. This last result shows how, at least in the present framework, the
Cosmic Mach Number is pretty insensitive to the underlying statistics (Gaussian
or skewed) of models with otherwise identical cosmological parameters.

\begin{figure}
\centering
\vspace{9cm}
\caption{Bulk flow $v_{bulk}$ (left panel), one--point velocity dispersion
$\sigma_v$ (central panel) and Cosmic Mach Number $\cal M$ (right panel) versus
the radius $R$. The models are represented as in Figure 2.}
\end{figure}

\section{Comparison with Observational Data}

\subsection{Sample and catalogue construction}

In this section we use the same observational data considered in Tormen et al.
(1993), where more details can be found. Our catalogue is a compilation of
peculiar velocities from the ``Mark II" data, and contains 1184 galaxies,
including spirals, ellipticals and S0. In this way we assume that different
types of galaxies are tracers of the same velocity field; this assumption was
supported by Kolatt \& Dekel (1994), who recently analysed the velocity fields
traced separately by ellipticals and spirals. We reduced distance uncertainties
by grouping the galaxies following the rules in the original papers
(Lynden--Bell et al. 1988; Faber et al. 1989): at the end our sample consists
of 704 objects.

We used the results of our N--body simulations to construct mock catalogues.
Our analysis relies on the assumption that the galaxy peculiar velocities give
an unbiased signal of the actual velocity field. Because of this, our mock
catalogues should not be interpreted as ``galaxy catalogues", but just as an
appropriate mask applied to the full three dimensional velocity field in order
to mimic the observations. In this sense we are allowed to neglect velocity
bias effects (Carlberg, Couchman \& Thomas 1990; Couchman \& Carlberg 1992),
which are in any case believed to mostly affect small scale peculiar
velocities.

We located in each simulation 500 ``observers" in grid--points with features
similar to those of the Local Group (LG) (e.g. Gorski et al. 1989; Davis,
Strauss \& Yahil 1991; Strauss, Cen \& Ostriker 1993; Moscardini et al. 1994).
Three different requirements were imposed to each of these observers:

\noindent i) its peculiar velocity $v$ is in the range of the observed LG
motion, $v_{LG,obs} = 627 \pm 22$ km ${\rm s^{-1}}$ (Kogut et al. 1993);

\noindent ii) the local flow around the mock LG is quiet with a small local
`shear', ${\cal S} \equiv |{\bf v} - \langle {\bf v} \rangle|/| {\bf v}| <
0.2$, where $\langle {\bf v} \rangle$ is the average velocity of a sphere of
radius $R= 750$ km ${\rm s^{-1}}$ centred on the LG;

\noindent iii) the density contrast in the same sphere is in the range $-0.2
\le \delta \le 1.0$.

The reference frame was built from each LG position imposing that the velocity
vector of the central point had the CMB dipole direction ($l=276\pm 3^\circ$,
$b=30\pm 3^\circ$; Kogut et al. 1993), with the direction of the remaining axis
randomly selected. Next, we constructed our catalogues by collecting, for each
of the 704 positions of the observed objects, the closest particle in the
simulation and considering the radial component of its velocity with respect to
the LG position.

We then introduced in the simulated catalogues the galaxy distance errors
present in the real data, by perturbing each distance and radial peculiar
velocity with a Gaussian noise (e.g. Dekel, Bertschinger \& Faber 1990),
$r_{i,p} = r_i + \xi_i \Delta r_i $ and $u_{i,p} = u_i - \xi_i \Delta r_i +
\eta_i \sigma_f,$ where $\xi_i$ and $\eta_i$ are independent standard Gaussian
variables; $\Delta r_i$ is the estimated galaxy distance error and
$\sigma_f=200$ km ${\rm s^{-1}}$ is an estimate of the Hubble flow noise.

Finally, let us comment that the low resolution of the numerical code is not a
problem in the comparison with real data. In fact, the exclusion from the
catalogue of objects in virialized regions, the grouping technique discussed
above and the Hubble flow noise adopted in the model to calculate the peculiar
velocities correspond to an intrinsic  smoothing in the observed dataset on a
scale comparable to the numerical resolution.

\subsection{Results of statistical tests}

In our comparison between  the real sample and the mock catalogues, we will use
four different observables: the Local Group constraints, the bulk flow, the
Cosmic Mach Number and the velocity correlation function. For definitions and
assumptions adopted here we refer to Moscardini et al. (1994). The results for
the skewed models are shown in Figure 4 and in Table 2: they can be compared
with analogous results for the Gaussian model presented in Moscardini et al.
(1994), where that model was analysed in the frame of non--scale invariant CDM
models.

First, we estimated the capability of our models to reproduce the
characteristics of the observers (Local Groups), i.e. their velocity, local
shear and local density contrast, as previously described. In Table 2 we report
the percentage of grid--points that separately fulfil each constraint [${\cal
P}(v)$, ${\cal P}(\delta)$ and ${\cal P}({\cal S})$ for the velocity, density
and shear constraints respectively] and altogether ${\cal P}(LG)$,  both for
the skewed models and the Gaussian one. As in similar previous analyses, we are
not able to discriminate between the models using the constraint on the flow
quietness: the differences are very low and the constraint is poorly stringent.
When we consider the velocity distribution, the Gaussian model is preferred to
all skewed ones, even if the skew--negative are better than the skew--positive
ones. The behaviour is opposite when the models are compared by the density
contrast constraint: skew--negative models have a larger number of grid--points
in the observed range. The balancing of these results gives similar total
probability of reproducing all the LG characteristics at the same time for
Gaussian and skew--negative models, while the skew--positive ones have smaller
percentages. In any case, the differences among the considered models are not
so large as naively expected by considering different primordial distributions
of the gravitational potential. This is probably due to the fact that we are
considering models with the same normalisation, i.e. $b=1$. In Tormen at al.
(1993) and Moscardini et al. (1994) it was found that the total probability
${\cal P}(LG)$ turns out to be strongly dependent on the bias parameter but
almost independent of other parameters, such as the spectral index and the
density parameter $\Omega$.

In the upper panels of Figure 4 we plot the distributions of bulk flow
amplitudes calculated from our mock catalogues. The continuous vertical line
refers to the observed value: $v_{bulk}=306 \pm 72$ km ${\rm s^{-1}}$, with a
misalignment angle $\alpha = 54^\circ \pm 13^\circ$ with respect to the
direction of the CMB dipole. The distributions appear similar for different
models, even if skew--positive models show a slightly longer tail toward high
amplitude. We checked that a Maxwellian distribution provides a good fit of the
data, as shown by a Kolmogorov--Smirnov test for all the considered models.

In Table 2, we report the probability ${\cal P}(v_{b})$ that the simulated bulk
flows have amplitude in the interval $[v_{obs} - \sigma_{v obs},~v_{obs} +
\sigma_{v obs}]$ and misalignment angle $\alpha$ in the analogous interval.
These results show that the Gaussian model reproduces the real data more
frequently than the skewed ones.

The central panels of Figure 4 show the distributions for the Cosmic Mach
Number as calculated from our simulated catalogues, and compares them with the
value observed for our galaxy sample: ${\cal M}=0.24\pm0.06$ (showed by the
vertical line). In Table 2, we report the probability ${\cal P}({\cal M})$ that
the simulated Cosmic Mach Number is inside the observed interval. This
statistic turns out to be once more the less stringent one, but on its basis we
can conclude that the Gaussian model is preferred. The skew--positive models
seem to have a  larger probability for higher values of ${\cal M}$. Once again,
using a Kolmogorov--Smirnov test we find that the distributions for $\cal M$
are well fitted by a Maxwellian function.

As a last statistic, we considered the velocity correlation function and in
particular its linear integral $J_v$ from the origin to the maximum considered
pair separation, $R_{max}=5,000$ km ${\rm s^{-1}}$ (see Gorski et al. 1989;
Tormen et al. 1993). For our real catalogue we find $J_v/(100 {\rm ~km
{}~s}^{-1})^3=237.9\pm 61.5$. The lower panels of Figure 4 show the
distributions
of $J_v$ obtained from our simulated catalogues while the percentages ${\cal P}
(J_{v})$ of the simulated catalogues whose value of $J_v$ is less than one
standard deviation different from the observed one are reported in Table 2. The
behaviour of different non--Gaussian models is similar, but worse than the
Gaussian model in reproducing the observational data.

Finally, we performed a Maximum Likelihood analysis to compare all the models.
Calling $\vec C$ the random vector of the statistics used to constrain the
simulated Local Groups, $\vec C=(v_{LG}, {\cal S}, \delta)$, and $\vec S$ the
vector of all the other statistics, $\vec S=(v_{bulk}, \alpha, {\cal M}, J_v)$,
the joint distribution of $\vec C$ and $\vec S$, under the condition $\vec
C=\vec C_{obs}$, is ${\cal P}(\vec C_{obs},\vec S) = {\cal P}(\vec C_{obs})
{\cal P}(\vec S |\vec C_{obs})$. For a given model $H$, the likelihood function
reads ${\cal L}(H)={\cal P}(\vec C_{obs}|H) {\cal P}(\vec S_{obs}|\vec
C_{obs},H)$. The joint conditional likelihood ${\cal P}(\vec S_{obs}|\vec
C_{obs},H)$ of $v_{bulk}$, misalignment angle $\alpha$, Cosmic Mach Number
$\cal M$ and correlation integral $J_v$ has been computed by counting the
number of simulated catalogues that have, at the same time, $v_{bulk}$,
$\alpha$, $\cal M$ and $J_v$ consistent with the observed values, within the
quoted error bars. Table 2 reports, for all the considered models, the
resulting values for the joint likelihood ${\cal L}(H)$. Unlike our previous
works (Tormen et al. 1993; Moscardini et al. 1994), here we are not allowed to
use the relative likelihood in order to give confidence intervals to our
results, since the differences between our models are not of parametric kind.

Even if the presently available observational data appear to be not so
stringent in discriminating between the scenarios here considered, on the basis
of our analysis we find that the best model is the Gaussian one; however, the
difference with respect to the $\chi^2_n$ is not so large. On the contrary, the
skew--positive models have a lower likelihood (approximately one third of that
of the Gaussian model) and seem not to provide a good alternative to the
standard Gaussian scenario. These results are coherent with our previous
analyses of the same models, mostly based on the clustering properties: a
positive primordial skewness does not help in overcoming the difficulties of
the standard CDM scenario, while the skew--negative models cannot be excluded
by present data.

\begin{figure}
\centering
\vspace{15cm}
\caption{Probability distributions for the absolute value of the bulk flow
$v_{bulk}$ (top row), the Cosmic Mach Number $\cal M$ (central row) and the
correlation integral $J_v$ (bottom row), for Lognormal (left column) and
Chi--squared models (right column). Skew--negative models are shown by thin
solid lines, skew--positive ones by dotted lines. The same distributions are
plotted also for the Gaussian model (heavy solid line). The vertical lines
refer to the one $\sigma$ range obtained from our real catalogue.}
\end{figure}

\begin{table}
\centering
\caption{Local Group constraints and Likelihood functions.}
\begin{tabular}{lcccccccc}
Model & ${\cal P}(v)$ & ${\cal P}(\delta)$ & ${\cal P}({\cal S})$ &
${\cal P}(LG)$ & ${\cal P}(v_{b})$ &${\cal P}({\cal M})$ &
${\cal P}(J_{v})$ & ${\cal L}$ \\
$G$       & 6.41 & 37.26 & 73.90 & 1.73 & 13.2 & 46.6 & 24.4 & 0.076 \\
$LN_n$    & 3.70 & 59.32 & 84.60 & 1.64 & 10.6 & 37.4 & 17.0 & 0.059 \\
$LN_p$    & 2.48 & 51.18 & 86.41 & 0.99 &  5.8 & 30.4 & 16.2 & 0.022 \\
$\chi^2_n$& 4.84 & 55.15 & 79.38 & 2.08 &  7.6 & 32.4 & 13.6 & 0.037 \\
$\chi^2_p$& 2.98 & 40.37 & 80.48 & 1.05 &  7.8 & 33.8 & 19.0 & 0.025 \\
\end{tabular}
\end{table}

\section{Conclusions}

In this paper we have analysed the large--scale velocity field in the context
of non--Gaussian CDM models. Weinberg \& Cole (1992) partially analysed the
properties of peculiar velocities in their non--Gaussian models, assuming
scale--free initial conditions. Our work presents the first detailed study of
large--scale motions as resulting from primordial non--random phases in a CDM
scenario.

Unlike previous analyses on the same models, mostly devoted to the study of the
matter distribution and its clustering properties, in this work we found that
the sign of the primordial skewness of the density field provides a poor
discriminatory power. All our non--Gaussian models tend to have lower velocity
dispersion and bulk flow than the standard Gaussian CDM model. We interpret
this as due to the effect of primordial phase correlation which, in
skew--positive models, causes power to be transferred from large scales
(important for velocities) to small scales, whereas in skew--negative models it
causes a general slowing down of the growth of the power--spectrum at all
scales. This result is different to our earlier findings due to the different
normalization here applied, which is dictated by the {\em COBE} data, i.e.
$b\approx 1$.

In particular, this choice does not allow our skew--negative models to fully
develop their dynamical properties, discovered in previous studies, where it
was found that only after a lengthy evolution these models could achieve the
right slope for the correlation function, assumed to indicate the ``present
time". These non--Gaussian models, therefore, only experience moderate
non--linear evolution (as shown by the low value of $\sigma_\delta$ in Table
1), which makes their large--scale velocity field still sensitive to the
initial conditions (i.e. to the CDM power--spectrum) and only marginally
dependent, through mildly non--linear effects, on its primordial kurtosis and
on the skewness of the density fluctuations. This very fact implies that the
comparison of our non--Gaussian models with observational data does not clearly
permit to single out a best model: the standard Gaussian model is marginally
preferred by the maximum likelihood analysis.

Moreover, these results suggest that, contrary to naive expectations,
primordial non--random phases do not help in producing large--scale bulk
motions such as those indicated by the Lauer \& Postman (1994) analysis, which
is one of the most challenging observational results for the present structure
formation scenarios.

\section* {Acknowledgments}

We thank David Burstein for providing us with the Mark II catalogue. This work
has been partially supported by Italian MURST and by CNR (Progetto Finalizzato:
Sistemi Informatici e Calcolo Parallelo). We acknowledge the staff and the
management of the CINECA Computer Centre (Bologna) for their assistance and for
allowing the use of computational facilities.

\end{document}